\documentclass[sigconf]{acmart}
\AtBeginDocument{%
  \providecommand\BibTeX{{%
    \normalfont B\kern-0.5em{\scshape i\kern-0.25em b}\kern-0.8em\TeX}}}

\copyrightyear{2021}
\acmYear{2021}
\setcopyright{acmcopyright}

\acmConference[SIGIR '21] {Proceedings of the 44th International ACM SIGIR Conference on Research and Development in Information Retrieval}{July 11--15, 2021}{Virtual Event, Canada.}
\acmBooktitle{Proceedings of the 44th International ACM SIGIR Conference on Research and Development in Information Retrieval (SIGIR '21), July 11--15, 2021, Virtual Event, Canada}
\acmPrice{15.00}
\acmDOI{10.1145/3404835.3463048}
\acmISBN{978-1-4503-8037-9/21/07}

\usepackage{booktabs}
\usepackage{multirow}
\usepackage{adjustbox}
\usepackage{tabularx}
\usepackage{array}
\usepackage{graphicx,subcaption}
\usepackage{enumitem}
\usepackage{savefnmark}
\usepackage{footnote}
\usepackage{float, amsmath, blindtext}
\usepackage[belowskip=0pt,aboveskip=0pt]{caption}
\usepackage{graphicx}
\usepackage{natbib}
\usepackage{arydshln}
\usepackage{tikz}
\usepackage{dsfont}
\usepackage{balance}

\newcommand{\tsc}[1]{\textsuperscript{#1}} 

\begin{document}
\fancyhead{}

\title{Text-to-Text Multi-view Learning for Passage Re-ranking}


\author{Jia-Huei Ju,\tsc{1} Jheng-Hong Yang,\tsc{2} and Chuan-Ju Wang\tsc{1}}
\affiliation{
  \institution{
  \tsc{1} Research Center for Information Technology Innovation, Academia Sinica,\\
  \tsc{2} David R. Cheriton School of Computer Science, University of Waterloo}
  \country{}
}

\renewcommand{\authors}{Jia-Huei Ju, Jheng-Hong Yang, Chuan-Ju Wang}

\renewcommand{\shortauthors}{Ju and Yang, et al.}

\begin{abstract}
Recently, much progress in natural language processing has been driven by deep contextualized representations pretrained on large corpora.  
Typically, the fine-tuning on these pretrained models for a specific downstream task is based on \textit{single-view} learning, which is however inadequate as a sentence can be interpreted differently from different perspectives. 
Therefore, in this work, we propose a text-to-text \textit{multi-view} learning framework by incorporating an additional view---the text generation view---into a typical single-view passage ranking model.
Empirically, the proposed approach is of help to the ranking performance compared to its single-view counterpart. 
Component analysis is also reported in the paper.
\end{abstract}


\maketitle
\section{Introduction}
Over the past few years, there has been increased interest in leveraging neural networks to facilitate text ranking in the information retrieval (IR) community~\cite{nogueira2020passage, Guo_2016, hof2020interpretable, mitra2017learning, Xiong_2017}.
Borrowing the great progress made by deep contextualized representations pretrained on large corpora~\cite{raffel2019exploring, devlin2019bert, radford2019language, brown2020language,lewis-etal-2020-bart} and the availability of large-scale human-annotated query-document pairs~\cite{bajaj2018ms}, text ranking models have experienced a great leap in ranking effectiveness compared to traditional IR baselines~\cite{lin2020pretrained}.
However, the great success of the text ranking model heavily relied on the number and the quality of training relevance pairs~\cite{zhang-etal-2020-little}.
Therefore, the urge to develop more data-efficient approaches is strong due to the expensiveness of collecting high quality human-annotated relevant pairs.

Recently, the success of deep contextualized representations, e.g., BERT~\cite{devlin2019bert}, is further advanced by the unified text-to-text pretrained transformer frameworks~\cite{raffel2019exploring, lewis-etal-2020-bart, brown2020language}.
By unifying natural language processing tasks in one text-to-text interpretation, these text-to-text pretrained models can be adopted for advancing existing IR applications in a straightforward manner such as passage and document ranking~\cite{nogueira2020document}, document expansion~\cite{nogueira2019doc2query}, and relevance data augmentation~\cite{bird2020chatbot}.
However, while these pretrained text-to-text models have further advanced ranking effectiveness from different angles, there still exist several shortcomings.
One of the challenges posed by leveraging pretrained text-to-text models is the overfitting issue, to alleviate which, the recent work~\cite{nogueira2020document} scales the number of parameters up to the formidable three billion.
Normally, the generalization ability can be amended by (1) pretraining models using more parameters on larger corpus, or (2) collecting more relevance labels of the text ranking task; in contrast to the previous two approaches, in this study, we aim at addressing the problem from the perspective of multi-view learning.
By taking a deeper look into the existing text-ranking and document expansion models, we hypothesize that combining multi-view learning with the text-to-text framework is a possible direction toward better contextualized representation for text ranking.
Mainly developed by the computer vision community, multi-view learning is a framework that seeks to learn better object representations via representation alignment and representation fusion~\cite{Li_2019}.
For example, an object's representation is more comprehensive when we train models to unify different views in 3D object recognition~\cite{su2015multiview} or multi-modal representation learning~\cite{ngiam2011multimodal}.
Inspired by these works, we conjecture that the concept of relevant pairs can be represented from two different \emph{views}: (1) a text-ranking objective; (2) a text-generation objective.
In addition, the text-to-text framework is by nature matched with the multi-view learning as it is easy to append different task prefix heads given the same relevance pair, i.e., a query and a relevant passage.

\begin{figure}[t!]
    \centering
    \includegraphics[width=0.9\linewidth]{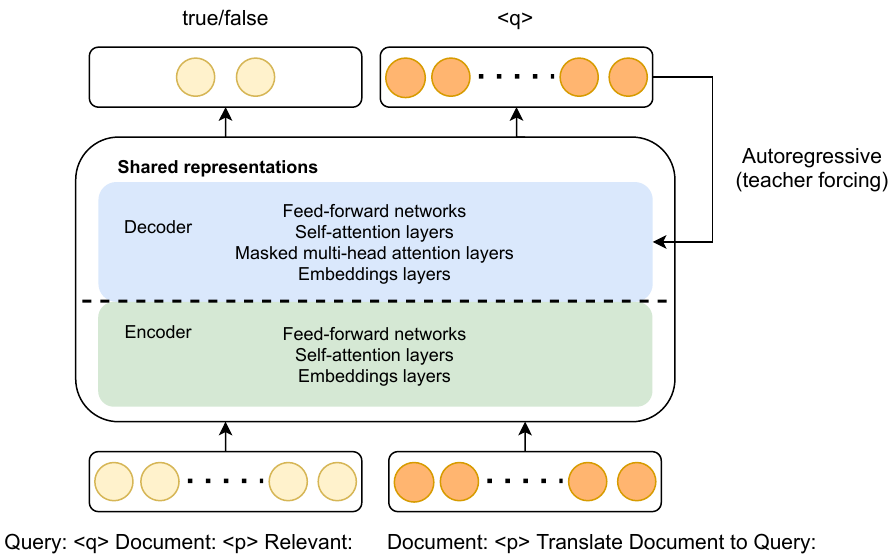}
    \caption{Text-to-text multi-view learning for the shared representations using the two objectives of passage ranking (left half) and text generation (right half).
    }
    \label{fig:arch}\vspace{-0.5cm}
\end{figure}

In sum, as illustrated in Figure~\ref{fig:arch}, we seek to combine the two popular objectives in IR tasks---passage ranking and text generation---to explore the opportunity of getting a better contextualized representation for text ranking with neither producing extra relevance pairs nor pretraining a larger model.
In the following sections, we first review the two popular applications of the text-to-text pretrained models developed by the IR community and introduce our multi-view learning approach: mixed instance sampling from different views.
Afterwards, we empirically examine our approach on the well-known large-scale text-ranking dataset: MS MARCO~\cite{bajaj2018ms}.
Finally, we report our observations from our experiments and component analysis, the summary of which is listed as follows.
\begin{itemize}[leftmargin=*]
    \item The mixing rate for instance sampling plays a vital role.
    \item The multi-view learning provides a more robust representation for passage re-ranking when increasing the number of candidates.
    \item The objective of text generation is complementary to passage ranking, which is sensitive to the prefix heads and source (target) placements of text-to-text formulations.
\end{itemize}

\section{Methodology}
In this section, we introduce the proposed multi-view learning approach with a text-to-text framework for passage ranking.
Specifically, our work is built on top of the text-to-text transfer transformers (T5)~\cite{raffel2019exploring} with a passage ranking model~\cite{nogueira2020document} and a query prediction model~\cite{nogueira2019doc2query}.
It is worth noting that, although we testify our ideology with a passage ranking scenario, our approach could be extended to document ranking with proper modifications.

\smallskip\noindent
\textbf{Passage ranking.}
The goal of a passage ranking model is to estimate the relevance score given a query $q$ and a candidate passage $p$.  
Following the work of~\cite{nogueira2020document}, which leverages the pretrained T5 model, we compute the relevance score of each pair ($q$, $p$) with the softmax normalized probabilities among two predefined tokens (\texttt{true} or \texttt{false}) conditioned on ($q$, $p$).
Our text-to-text model that infers the conditional probability is then fine-tuned with the following negative log likelihood loss:
\begin{equation*}
\resizebox{0.91\hsize}{!}{
$
\mathcal{L}_\text{Rank}(q, p^+, p^-) =  
-\log P(\texttt{true}\ | q, p^+ )
- \log P(\texttt{false}\ | q, p^- ),
$
}
\end{equation*}
where a training input triplet is composed of a query $q$, a relevant passage $p^+$, and a non-relevant passage $p^-$.

\smallskip\noindent
\textbf{Text generation.}
Query prediction is a typical text generation task that produces texts (query) conditioned on input texts (passage); we here term it as a passage-to-query task (P2Q, hereafter). 
The previous work in~\cite{nogueira2019doc2query} fine-tunes the T5 model on labelled passage-query pairs to obtain a P2Q model, by which one can enhance the effectiveness of passage ranking by document expansion using predicted queries~\cite{nogueira2019document} or data augmentation using the inferred relevance (or so-called weakly supervised) pairs on the unlabelled data~\cite{ma2021zeroshot}.
Our objective for training a typical P2Q model is:
\begin{equation*}
\mathcal{L}_\text{P2Q}(q, p)
= -\sum_{t=1}^{|q|}\log P(q_{(t:t)}\ | q_{(1:t-1)}, p),
\end{equation*}
where $|q|$ denotes the length of query $q$ and $q_{(j:k)}$ represents the sub-query extracted from $q$ beginning at the $j$-th word and extending to the $k$-th word.

\smallskip\noindent
\textbf{Multi-view learning.}
Inspired by the multi-view learning~\cite{tang2019improving, ngiam2011multimodal}, in this paper, we build a unified approach based on the T5 model that simultaneously considers two views, a primary Rank view and a auxiliary P2Q view, for the task of passage ranking.
To be more specific, different from previous two-step approaches that append the predicted queries to the document to enhance the effectiveness of passage ranking (e.g.,\cite{nogueira2019document}), we propose to jointly train a shared representation based on the concept of the model-level fusion~\cite{d2015review} in the multi-view learning.
In this paper, we hypothesize that the shared representation has better generalization ability when we fuse the objectives of the tasks of passage ranking and text generation, for which the objective is defined as
\begin{equation*}
\mathcal{L}_\text{multi-view}=(1-X)\times \mathcal{L}_{\text{Rank}}(q, p^+, p^-)+X\times \mathcal{L}_{\text{P2Q}}(q, p),
\end{equation*}
where $X\sim{\rm Bernoulli}(\eta)$\footnote{We here adopt the multinomial sampling method provided in~\url{https://github.com/google-research/text-to-text-transfer-transformer}.} and $\eta$ is a predefined parameter termed as the ``mixing rate'' hereafter.
Note that in the sense of the numbers of training examples from the two views, the mixing rate $\eta$ equals to the proportion of the examples of text generation view to the total number of examples from the two views. 

\section{Empirical Evaluation}
\subsection{MS MARCO Passage Ranking}
We validate our method on the MS MARCO passage ranking dataset (MARCO, henceforth)~\cite{bajaj2018ms} which contains 8.8M passages.
The models are trained on two types of data: 
(1) \texttt{triples.train.small} is the official training triples composed of queries $q$, positive passages $p^{+}$, and negative passages $p^{-}$; (2) \texttt{qrels.train} contains the relevant query--passage pairs with their ids ($q^{*}$, $p^{*}$), respectively.
The former one, (1), is used for passage ranking training, while the latter one, (2) is used for text generation training and component analysis.

For evaluation, we use the sparsely judged MARCO query set of 6,980 queries as our development set (abbreviated as Dev hereafter) to select hyperparameters and conduct component analysis.
To verify the generalization capability of our method, we report our final results with a larger testing set. We take the rest of the full sparsely judged MARCO query set of 51,836 queries (abbreviated as Dev-Rest) for testing.
The evaluation metric is MRR@10, which is aligned with the official leaderboard.

\smallskip\noindent
\textbf{Two-stage passage ranking.}
We use a standard two-stage passage ranking pipeline to facilitate the passage ranking tasks in the following experiments.
For the first-stage passage filtering model, we adopt BM25 with a fixed hyperparameter set to $(k_1=0.82, b=0.68)$.
Then, we feed the queries together with their filtered passage candidates into the second-stage passage re-ranking model to produce the final ranking list.

\smallskip\noindent
\textbf{Text-to-text backbones.}
Following the work of~\cite{nogueira2020document}, we adopt the same process to prepare training data and fine-tune the T5 model for passage re-ranking using the checkpoints of three sizes: T5-base, T5-large, and T5-3B.\footnote{We align our hyperparameters with the ones in~\cite{nogueira2020document}, where we adopt adafactor with constant learning rate = $10^{-3}$, batch size = 128 and train our T5 models for 100K steps.}
In addition, given a query-passage pair, we adopt the logit trick proposed by ~\cite{nogueira2020document} to inference relevance scores, which are softmax normalized probabilities among two predefined tokens (case sensitive): \texttt{true} and \texttt{false}.
As for the text generation task, we follow~\cite{nogueira2019document} to prepare our formulations of different views; we then apply the vanilla teacher forcing to supervise the T5 model for text generation.
Figure~\ref{fig:arch} illustrates our text-to-text formulations for different views (left half and right half).

\smallskip\noindent
\textbf{Multi-view learning.}
We jointly train the T5 models with the passage ranking task ({\bf Rank} view) and the text generation task ({\bf P2Q} view) with the standard example-proportional mixing technique~\cite{raffel2019exploring}.
In our study, we observe that the mixing rate plays a vital role in our multi-view learning framework, which is similar to the findings in the earlier work of multi-task learning~\cite{arivazhagan2019massively}.
Hence, we first search for the best mixing rate with the performance on Dev set with T5-large and adopt the best mixing rate $\eta=0.15$ to evaluate the overall performance (see Section~\ref{sec:mix} for more details).
Note that in the following experiments, we use the same mixing rate for T5-base and T5-3B as searching hyperparameters for them is infeasible due to our limited computation resources.

\subsection{Main Results}
\label{sec:main}
In the experiments, for each of T5-base, T5-large, and T5-3B models, we conduct the fine-tuning ten times with different random seeds.
Afterward, we take Dev as our validation set and select the \textit{best} fine-tuned models among the ten models by measuring their ranking effectiveness (see Section~\ref{sec:mix} for more detail).
Furthermore, we test our model on Dev-Rest to verify the generalization capability of the proposed multi-view learning framework.

\smallskip\noindent
\textbf{Re-ranking effectiveness.}
We first attest the overall re-ranking effectiveness of the multi-view learning, the results of which are tabulated in Table~\ref{tb:main_result}.
In the table, we highlight the scores of the multi-view condition in boldface when it is significantly better than its single-view counterpart with a paired $t$-test $(p\leq0.05)$.
The first observation is that the multi-view re-ranking model has generally better ranking effectiveness; specifically, the MRR@10 scores on the Dev set are increased by 0.001 (T5-base), 0.006 (T5-large), and 0.004 (T5-3B), respectively. 
Particularly, it is worth noting that the T5-large (condition 5) is on par with T5-3B (condition 3) in terms of ranking effectiveness, but with much fewer parameters.
Moreover, observed from the last column (Dev-Rest) in Table~\ref{tb:main_result}, there is still positive impact brought from the proposed multi-view learning; among all models, T5-large obtains the greatest performance improvement in terms of MRR@10.


\begin{table}
    \centering
    \resizebox{\columnwidth}{!}{
    \begin{tabular}{lllcccc}
    \toprule
        \# & Condition & Model & \# Param (M) &
        Dev & Dev-Rest \\
    \toprule
        & \multirow{3}{*}{\textbf{Baselines}} & BM25   & -  & 0.187  & 0.191 \\
        & & Best non-BERT~\cite{hof2019effect}     & - &  0.290 &  -  \\
        & & BM25 + BERT-large~\cite{nogueira2019multistage} & 340 &  0.372 &  -  \\
    \midrule
        1 & \multirow{3}{*}{\textbf{Single-view}} & BM25 +T5-base  & 220  & 0.384  &  0.380  \\
        2 & & BM25 +T5-large & 770  & 0.395  &  0.390  \\
        3 & & BM25 +T5-3B    & 2,800 & 0.398  &  0.395  \\
    \midrule
        4 & \multirow{3}{*}{\textbf{Multi-view}} & BM25 +T5-base  & 220  & 0.385  & \textbf{0.382}$^{1}$  \\
        5 & & BM25 +T5-large & 770  & \textbf{0.401}$^{2}$ & \textbf{0.393}$^{2}$ \\
        6 & & BM25 +T5-3B    & 2,800 & 0.402  & 0.396   \\
    \bottomrule
    \end{tabular}
    }
    \caption{Comparison on overall ranking effectiveness (MRR@10).
    The scores are in boldface if they are significantly better than the compared condition (see the superscript) under a paired $t$-test with $p \leq 0.05$.}
    \label{tb:main_result}
    \vspace{-0.5cm}
\end{table}

\begin{figure}[ht!]\vspace{-0.4cm}
    \centering
    \includegraphics[width=0.95\linewidth]{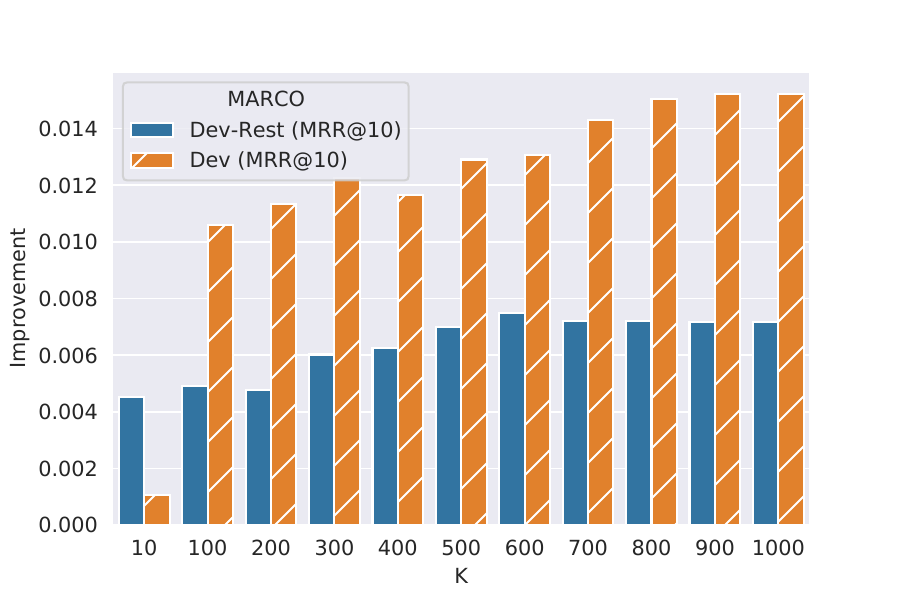}
    \caption{Improvement of MRR@10 with top-$K$ candidates based on the BM25.
    The re-ranking model is T5-large (multi-view versus single-view).
    }
    \label{fig:topk-imprv}
    \vspace{-0.4cm}
\end{figure}

\smallskip\noindent
\textbf{Re-ranking effectiveness at different depths.}
To better understand the advantages of the multi-view learning, we design a sweeping depth experiment to testify the ranking robustness regarding the number of candidates.
We conduct the re-ranking task by sweeping the depth of the BM25 retrieved list; we can therefore validate the marginal effect of noises in different degrees by cutting off the retrieved list with $K$ candidates based on BM25 scores.
Specifically, we define:
$\text{Improvement} = \frac{\text{MRR}@10_\text{multi} - \text{MRR}@10_\text{single}}{\text{MRR}@10_\text{single}}$ and evaluate T5-large as the re-ranking model on Dev and Dev-Rest.
Figure~\ref{fig:topk-imprv} illustrates the results for a range of $K$, where the $x$-axis indicates the number of candidates $K$, and the $y$-axis represents the improvements defined above.

Overall, the proposed multi-view learning improves the ranking effectiveness within the depth of 1000.
Additionally, we observe that the improvement gradually increases when the depth is swept from 10 to 1000.
Although the ranking robustness against the noisy environment is decreased when we generalize it from Dev to Dev-Rest, there is still observable improvement increasing w.r.t. $K$.
This observation implies that the multi-view learning helps the re-ranking models to discriminate the relevant passage in the more noisy environment (larger $K$).

\begin{figure}[H]\vspace{-0.4cm}
    \centering
    \includegraphics[width=0.95\linewidth]{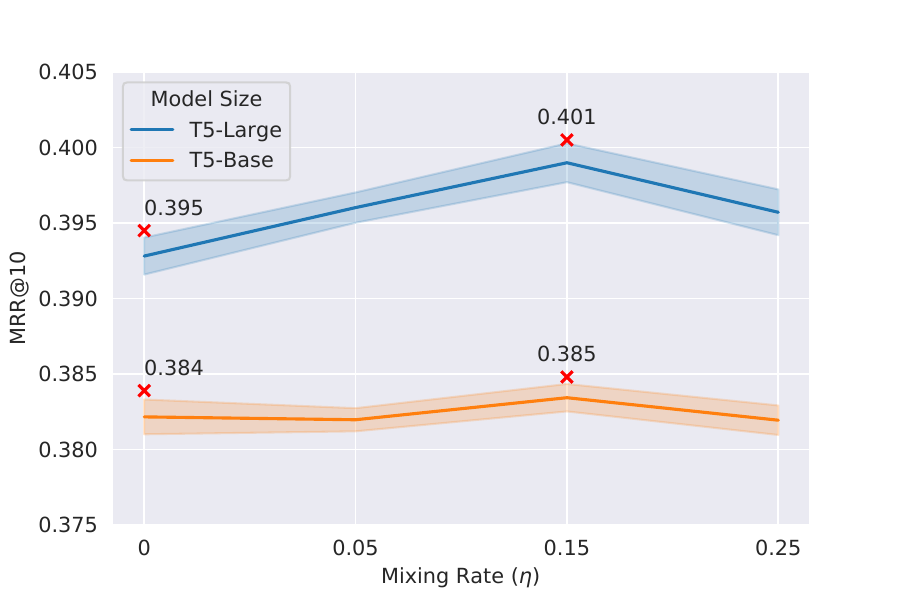}
    \caption{Impact of the mixing rate ($\eta$) on T5-base and T5-large models. We report the MRR@10 on Dev, including the mean values and corresponding 95\% confidence interval of the ten models and mark the \textit{highest} value with crosses.}
    \label{fig:mixingrate}
    \vspace{-0.4cm}
\end{figure}

\subsection{Impact of Mixing Rates}
\label{sec:mix}
From the experiments, we observe that the mixing proportion of the passage ranking task ({\bf Rank} view) and the text generation task ({\bf P2Q} view) has a notable impact on the effectiveness of our passage re-ranking model.
To investigate the impact in depth, we fix other hyperparameters and conduct experiments with various $\eta$.
Recall that the mixing rate ($\eta$) is equal to the proportion of the examples of text generation view to the total number of examples; for example, $\eta=0.15$ denotes that our total training examples comprise $15\%$ {\bf P2Q} and $85\%$ {\bf Rank} views.
Considering the randomness in the fine-tuning process, we conduct ten experiments for each $\eta$ and report the mean and the 95\% confidence interval of MRR@10 measured in the Dev set.
Due to the limitation of our computation resources, we here conduct the experiments on T5-base and T5-large models only.
Figure~\ref{fig:mixingrate} illustrates the results for the mixing rate $\eta = \{0, 0.05, 0.15, 0.25\}$, where the crosses denote the \textit{best} scores among the ten models.

The first observation from Figure~\ref{fig:mixingrate} is that the ranking effectiveness on T5-large follows a hump-shaped pattern, with a peak of $\eta=0.15$.
Second, there is no explicit improvement on T5-base as the improvements from $\eta = 0 \sim 0.15$ are mostly within the range of the 95\% confidence interval at $\eta = 0$.
With the above results, we conjecture that the multi-view learning is limited by the model capacity since training a smaller model on more data is often outperformed by training a larger model for fewer steps~\cite{raffel2019exploring}.

\section{Component Analysis}
\label{sec:abl}
In this section, we conduct detailed experiments to dissect the proposed text-to-text multi-view learning strategy.
Starting by the single-view learning as our baseline, we investigate the research questions from two major perspectives: template and prefix, and dive into their view variants in terms of the text-to-text formulations in Table~\ref{tb:formulation}.
In the following experiments, we adopt T5-large as our primary subject and conduct five fine-tuning processes ($\eta=0.15$) for each view combination.\footnote{Note that the reason for only conducting five experiments for each view combination is that we observe that the score variations here are relatively small compared to the mixing rate experiments in Section~\ref{sec:mix}.}
The results are reported in Table~\ref{tb:ablation}, where the means ($\pm$ standard deviation) of MRR@10 measured in the Dev set are tabulated.

\begin{table}
    \centering
    \resizebox{\columnwidth}{!}{
    \begin{tabular}{lcc}
    \toprule
        View & Source & Target \\
    \midrule
        \multirow{2}{*}{\textbf{Rank~\cite{nogueira2020document}}} 
        & Query: <$q$> Document: <$p^{+}$> Relevant: & true \\
        & Query: <$q$> Document: <$p^{-}$> Relevant: & false \\
    \midrule
        \multirow{2}{*}{\textbf{Rank (swap)}} & Relevant: Query: <$q$> Document: <$p^{+}$> & true \\
        & Relevant: Query: <$q$> Document: <$p^{-}$> & false \\
    \bottomrule
    \toprule
        \textbf{P2Q} & Document: <$p^{*}$> Translate Document to Query: & <$q^{*}$> \\
    \midrule
        \textbf{P2Q (swap)} & Translate Document to Query: Document: <$p^{*}$> & <$q^{*}$> \\
    \midrule
        \textbf{Q2P} & Query: <$q^{*}$> Translate Query to Document: & <$p^{*}$> \\
    \midrule
        \textbf{Rank}$^{*}$ & Query: <$q^{*}$> Document: <$p^{*}$> Relevant: & true \\
    \bottomrule
    \end{tabular}
    }
    \caption{Text-to-text formulations for View-1 and View-2.
    We put the query text $q$ and passage text $p^{(\cdot)}$ retrieved from our training data in the placeholders <$\cdot$> of these formulations.
    Note that our formulations are case-sensitive.}
    \label{tb:formulation}
    \vspace{-0.5cm}
\end{table}

\smallskip\noindent
\textbf{Template impact.}
First, we study the impact on the differences of the text-to-text templates by changing the formulations of the second view (View-2) while fixing the first view (View-1) as {\bf Rank}, shown in the second group (Condition: Template) of Table~\ref{tb:ablation}.
As shown in Table~\ref{tb:formulation}, we take queries and passages from \texttt{qrels.train} to construct the data for View-2, so we can dissect the impact from different text-to-text formulations while keeping as many factors fixed as possible.
In sum, we have two observations as follows:
\begin{itemize}[leftmargin=*]
    \item \textit{Formulation matters.}
    Comparing the entries of the single-view baseline and multi-view with Rank$^{*}$ as View-2, 
    we observe that the multi-view learning deteriorates the ranking effectiveness from 0.393 to 0.391.
    On the other hand, given the same amount of training data, the text generation views (P2Q and Q2P) outperform the baseline by 0.399 and 0.394.
    \item \textit{Source (target) placement matters.}
    Given the same text generation formulation, the source (target) placements of queries and passages also affect the ranking effectiveness.
    The score of P2Q (0.399) is higher than its counterpart, a reverse source-target placement, Q2P (0.394).
    We hypothesize that the difference is from the asymmetric properties of task difficulties since Q2P requires T5 to infer longer texts of passages from shorter texts of queries.
\end{itemize}

\smallskip\noindent
\textbf{Prefix dependency.} 
Task-specific prefix heads play a vital role in the text-to-text framework by nature~\cite{raffel2019exploring,schick2021exploiting,schick2020its}.
In this ablation study, we explore the impact of prefix heads' position dependencies from the source side. 
To be more specific, we move the Rank view's head \texttt{Relevant:} from the end to the beginning of a sentence, and do the same movement for the head of P2Q, \texttt{Translate Document\dots}, which are abbreviated as Rank (swap) and P2Q (swap) in Table~\ref{tb:formulation}.
In Table~\ref{tb:ablation}, we observe that the positional adjustments only matter on the Rank view.

The difference between Rank and P2Q is that Rank requires T5 to generate a fix-length target for every sentence, while P2Q requires T5 to generate target queries in different lengths.
Hence, the source position dependency may be more important for Rank, as the P2Q relies on both the source and the target.

\begin{table}
    \centering
    \small
    \begin{tabular}{lllc}
    \toprule
        Condition & View-1 & View-2 ($\eta=0.15$) & MRR@10 \\
    \midrule
        Baseline & Rank  & -          & 0.393 ($\pm0.001$)  \\
    \midrule
        \multirow{3}{*}{Template}
         & Rank  & Rank$^{*}$ & 0.391 ($\pm0.001$)  \\
         & \textbf{Rank}  & \textbf{P2Q}        & \textbf{0.399} ($\pm0.001$)  \\
         & Rank  & Q2P        & 0.394 ($\pm0.002$)  \\
    \midrule
        \multirow{3}{*}{Prefix}
           & Rank  & P2Q (swap)       & 0.399 ($\pm0.001$)  \\
           & Rank (swap) & P2Q        & 0.394 ($\pm0.001$)  \\ 
           & Rank (swap) & P2Q (swap) & 0.395 ($\pm0.001$)  \\
    \bottomrule
    \end{tabular}
    \caption{Component analysis of ranking effectiveness with different text-to-text formulations.}
    \label{tb:ablation}
    \vspace{-0.5cm}
\end{table}

\section{Conclusion}
We introduce the idea of multi-view learning into the existing text-to-text passage re-ranking model. 
Through the proposed text-to-text multi-view framework, we fuse the text-generation objective with the text-ranking objective by the instance mixing approach.
In our empirical study, we observe that the text generation view is beneficial in advancing re-ranking effectiveness.
Moreover, the results suggest that the most important factor is the mixing rate for sampling instances from different views.
Furthermore, we verify the multi-view model's re-ranking robustness via increasing its re-ranking depth.

Even though the connections between different views are still ambiguous, we consider multi-view learning as a flexible framework to achieve a better generalized representation with simple extensions.
For future work, possible directions include: (1) incorporating query prediction given non-relevant pairs; (2) fusing graded relevance scores of the existing term-matching models such as BM25.
Finally, we also consider exploring the connections underlying different views and how do they contribute to ranking robustness.

\section*{ACKNOWLEDGEMENTS}
We would like to thank the support of Cloud
TPUs from Google’s TPU Research Cloud (TRC).
\bibliographystyle{ACM-Reference-Format}
\balance
\bibliography{paper.bib}

\end{document}